\documentclass[12pt]{article}

\usepackage{hyperref}\usepackage{bookmark}

\newcommand{\Sect}[1]{Sect.~\ref{#1}}
\newcommand{\bra}[1]{{\langle #1|}}
\newcommand{\ket}[1]{{|#1\rangle}}
\newcommand{\eq}[1]{(\ref{#1})}
\newcommand{\Eq}{Eq.~\eq}
\newcommand{\proj}[1]{{\ket{#1}\bra{#1}}}

\title{Super-intuition and correlations with the future in Quantum Consciousness}
\author{Michael B. Mensky\\
{\small P.N. Lebedev Physics Institute,} 
{\small Russian Academy of Sciences,} \\
{\small 53 Leninsky prosp., 119991 Moscow, Russia}\\
{\small Email: mensky@lpi.ru}}
\date{}

\begin{document}

\maketitle

\tableofcontents

\newpage

\begin{abstract}	
The role of quantum information is discussed in the framework of Quantum Concept of Consciousness (QCC), based on the many-worlds interpretation of quantum mechanics (Everett interpretation). Within QCC the phenomenon of super-intuition is analyzed, which explains in particular the great scientific insights and realizes a sort of the ``mental time machine''. The recently expressed opinion that emergence of super-intuition requires transferring quantum information, which is banned by the impossibility of ``quantum cloning'', is critically considered. It is shown that in fact under QCC the emergence of the phenomenon of super-intuition requires not transferring quantum information, but only correlating various fragments of it. Actual examples of scientific insights that have been documented by prominent scientists, suggest that the marker of the correlation in this case is a strong positive emotion, which can be called cognitive euphoria.
\end{abstract}

\newpage


\section{Introduction}
\label{sec:Intro}

Quantum mechanics has at its origin put a number of conceptual problems \cite{Bohr49, NeumannENG55bk, WheelerZur83collect, Wigner83mind, Bell64}, which are not resolved until now. These problems are known under the term of \emph{measurement problem}  since are revealed in the consideration of the measurement procedure within the formalism of quantum mechanics. Indeed, this formalism leads to the paradoxical conclusion, that the states of the measuring apparatus, which correspond to different measurement readouts, although differ on a macroscopic scale, must coexist (from a formal point of view---be in a superposition).

The most common so-called \emph{Copenhagen interpretation} of quantum mechanics introduces the reduction postulate, according to which a superposition of macroscopically distinct states disappears, and only one component of this superposition remains. However, the reduction postulate actually contradicts the formalism of quantum mechanics and therefore it violates the logic of the whole theory.

In the middle of 20th century, American physicist \emph{Hugh Everett proposed a different interpretation of quantum mechanics} \cite{Everett57, DeWittGrah73everett, Deutsch1997engBk, Men00consEn, Grib-Everett-2013-ENG}
according to which macroscopically distinguishable states really may coexist (as terms of a superposition), which is a qualitative difference between the \emph{quantum concept of reality} from the concept of reality in the classic theory. Quantum reality is described in Everett interpretation as the \emph{coexistence of distinct classical realities} (Everett's worlds), which from the classical point of view are mutually exclusive. For this reason, the interpretation of Everett's is often called \emph{many-worlds interpretation}.\footnote{However, this term is in fact misleading, see \Sect{sec:EverettCoexostenceRealities}.}

The present author has proposed \emph{Quantum Concept of Consciousness} (QCC), or \emph{Extended Everett Concept} \cite {Men00consEn, Men05consEn, MBM-QCC-2007-ENG, MBM-QCC-postcorrection-2007,MBMConsInJOC2011}, making use of Everett's interpretation for a better understanding of \emph {consciousness and super-consciousness} (similar to \emph{the collective unconscious} of Carl Jung). According to QCC, besides the usual conscious perception of reality, when only one classical reality (from those superposed as components of the quantum reality) is available, under certain conditions may occur \emph {super-consciousness}, having access to the whole of quantum reality, that is, to all coexisting classical realities. The most complete exposition of the concept, see in \cite{MBMcons2010bk-EN}, while the current discussion of the various aspects of it --- in  \cite{MBM-QCC-2013, Zeh-QCC-2013, Panov-QCC-2013, Namiot-QCC-2013}.

Access of the super-consciousness to all alternatives (alternative classical realities) allows \emph{super-intuition}, or direct vision of the truth, the ability which is inaccessible for consciousness without aid of super-consciousness. In particular, super-intuition has access to the information which will be acquired in the future, but which does not yet exist in the present (see below in section \ref{sec:QCCandSuper-intuition}).

In the paper \cite{Panov-QCC-2013} A.D.~Panov points to a difficulty which the hypothesis of super-intuition may face. In the verbal statement of the hypothesis of super-consciousness, it has been stated that super-consciousness has access to information from all alternatives, including those relating to the future, and this information can be used in the present in a state of normal consciousness. Such a formulation can be understood as an assertion about the \emph{transfer of quantum information} from one space-time region to another.

The difficulty may result from theorem about impossibility of quantum cloning. This theorem prohibits in particular the transfer of quantum information from one space-time region to another without destroying the quantum information in the region where this information exists prior to the transfer. Thus, the phenomenon of super-intuition, as it seems, has to destroy the information in the future. It is shown in the paper \cite{Panov-QCC-2013} that this difficulty can be overcome if we assume that any quantum information is duplicated in the form of classical information.

In the present paper we will show that no additional assumptions are required, since this type of difficulty actually does not exist for the hypothesis of super-intuition in QCC. The reason is that the occurrence of super-intuition requires \emph{not transfer of quantum information, but only the detection of correlation} between fragments of quantum information in various areas of space-time.

Observation of real examples of super-intuition shows that the correlation arising in this phenomenon, is accompanied by a strong positive emotional tone, which can be called \emph{cognitive euphoria}. Cognitive euphoria plays a crucial role at key moments of scientific work, which relate to discovery qualitatively new approaches (paradigms) in the solution of scientific problems.

\section{Realities in Everett's Interpretation}
\label{sec:RealitiesInEverettInterpretation}

Quantum Concept of Consciousness has been initially formulated in the verbal form. However, it is evidently actual the task to expand, or adapt, the mathematical formalism of quantum mechanics in order to include QCC. Consider here the possible solution of this task (see also \cite{MBM-QCC-2013}).

\subsection{Measurements: Coexistence of macroscopically distinct states}
\label{sec:QuMeasAndCoexistenceMacroDistinctStates}

Quantum theory of measurement deals with restricted systems. Apart from the measured system $\ket{\psi}$, another system $\ket{\Phi}$ (the measuring device, meter, or measuring environment) is considered. Interaction of both systems is supposed to be arranged in a special way, so that, after a predetermined time interval (measurement time), the state of the measuring device can be used to judge about the state of the system being measured.

In the simplest case the meter distinguishes from each other the states $\ket{\psi_i}$, $i = 1, 2, \dots$ of the measured system, which have special properties with respect to the measurement (the basis formed by these states, is called ``pointer basis''). In terms of the total system including both measured subsystem and its environment (meter), the initial state of the form $\ket{\psi_i}\ket{\Phi_0}$ changes, after the measurement time, as follows:
$$
U\ket{\psi_i}\ket{\Phi_0}=\ket{\psi_i}\ket{\Phi_i}
$$
Here $\ket{\Phi_0}$ is a specially selected initial state of the device, $U$ --- evolution operator of the complete system during the measurement time (this system is assumed to be closed), and $\ket{\Phi_i}$ --- the state of the device, which indicates that the measured system is found to be in the corresponding state $\ket{\psi_i}$ of the pointer basis.

Due to linearity of the evolution operator $U$, for an arbitrary state of the measured system $\ket{\psi}=\sum_ic_i\ket{\psi_i}$, its evolution during the measurement is
\begin{equation}\label{MeasEvolution}
	U\sum_i c_i \ket{\psi_i}\ket{\Phi_0}=\sum_i c_i \ket{\psi_i}\ket{\Phi_i}
\end{equation}

This law generates known quantum paradoxes. The reason for this is that the states $\ket{\Phi_i}$, $i=1,2,\dots$, of the (macroscopic) device, should be macroscopically distinguishable. Therefore, the total system is after measuring in a superposition of macroscopically distinct states $\ket{\Psi_i}=\ket{\psi_i}\ket{\Phi_i}$, $i=1,2,\dots$. We can say that \emph{macroscopically distinguishable states should, in accordance with this law, coexist}. In the case of Schr\"{o}dinger's cat paradox (in which a cat plays essentially the role of the measuring device) the final state of the complete system includes a superposition of two states, one in which the cat is alive and the other with the dead cat. 

In other words, a quantum-mechanical analysis leads inevitably to the conclusion that a state  can (should) exist in which a cat is in the same time alive and dead. In a general form, the linear character of the quantum-mechanical evolution law guarantees that \emph{macroscopically distinct states may and even should coexist} (be components of superpositions). 

\subsection{Everett: Coexistence of distinct classical realities}
\label{sec:EverettCoexostenceRealities}

In the so-called \emph{Copenhagen interpretation} of quantum mechanics, it is assumed that after the measurement, the state of the complete system is not described by a superposition as in \Eq{MeasEvolution}, but with probability $p_i=|c_i|^2$ goes to one of the components of this superposition. This (in fact unjustified) transition is called \emph{reduction} of the quantum state, or \emph{collapse} of the wave function. This assumption is easily accepted by our intuition and is consistent with the results of all experiments. Therefore,  the Copenhagen interpretation can be accepted in probabilistic calculations (if probabilities of various results of measurements should be predicted). However, as the analysis in \Sect{sec:QuMeasAndCoexistenceMacroDistinctStates} shows, the assumption of state reduction is incompatible with the mathematical formalism of quantum mechanics (namely, the linear nature of the evolution of quantum systems). 

Therefore, reduction (and Copenhagen interpretation) cannot be accepted if one has to know \emph{what happens in reality}. The concepts of super-consciousness and super-intuition appearing in the framework of Quantum Concept of Consciousness (see \Sect{sec:QCCandSuper-intuition} below) show that this statement is not abstract (philosophical) but \emph{can be verified in the course of observing the work of consciousness and super-consciousness}. 

American physicist Hugh Everett was the first to suggest \cite{Everett57} that the mathematical formalism of quantum mechanics is more reliable guide than our intuition and assumed that the state of the form \eq{MeasEvolution} really arises in the measurement, i.e. a superposition of macroscopically distinct states inevitably emerges. This leads to the interpretation of quantum mechanics, which is called the \emph{Everett interpretation}.

To our intuition, which is educated by experience of observing macroscopic (i.e. classical) systems, it is really hard to accept this assumption, so it at least needs to be clarified. Renowned physicist Bryce DeWitt suggested \cite{DeWittGrah73everett} wording that helps to take Everett's interpretation. According to this verbal formulation, there is a set of classical worlds (Everett's worlds) after the measurement, and only one (say, $i$-th) measurement result (i.e. the state $\ket{\psi_i}\ket{\Phi_i}$ of the complete system) is realized in each of the worlds. 

It is then assumed that for each observer there is his ``twin'', or ``clone'', in each of Everett's worlds. With probability $p_i=|c_i|^2$ the observer identifies himself with the corresponding ($i$-th) of his twins. He sees, of course, only one ($i$-th) measurement result, and while sharing information with any other observer he deals with the $i$-th twin of this observer who sees the same ($i$-th) measurement result. Thus, each observer sees a classical world around him, and the observations of different observers are consistent.

The verbal description of the situation as coexisting many worlds has some psychological benefits and is usually accepted by adepts of Everett interpretation. Therefore, the latter received the name of \emph{many-worlds interpretation}. However, with this wording, one erroneous conclusion is often made, and this seems permanently slowed adoption of Everett interpretation by the scientific community. Let us explain this point.

When using the picture of many (Everett's) worlds, one has to say that the classical world, which has been observed before the measurement, is split in the measurement into many worlds, one for each measurement result. For those who are not sufficiently mastered the idea of  Everett's interpretation, this gives rise to rejecting the conception, because the conversion of one world into a number of worlds (usually into an infinite set of them) seems to be obviously impossible. Everett interpretation seems then to be completely unacceptable (they say, for example, that splitting of words results in energy non-conservation in this interpretation, which, of course, completely wrong).

To avoid this difficulty, it is much better to talk about \emph{many classical realities} rather than about many classical worlds. The set of many classical realities, in their totality, characterize \emph{a single quantum reality} of the material world, which is quantum in its nature \cite{Men00consEn}. No splitting of the world happens in the measurement. Just the state of our quantum world takes the form of \Eq{MeasEvolution}. This state describes quantum reality of the world after the measurement, which is presented by a set of classical realities (which are mutually incompatible from the classical point of view).

\section{Quantum Concept of Consciousness and super-intuition}
\label{sec:QCCandSuper-intuition}

Thus, according to Everett \cite{Everett57, DeWittGrah73everett}, the state of a quantum system is, in the general case, a sum of the state vectors corresponding to macroscopically distinct states. This can be formulated as a statement that \emph{quantum reality} is presented by a superposition of the state vectors corresponding to different \emph{classical realities} \cite{Men00consEn}. Simplifying, such a superposition can be represented as the sum 
\begin{equation}\label{Coexistence}
	\ket{\Psi} = \ket{\Psi_1} + \ket{\Psi_2} + \dots + \ket{\Psi_i} + \dots 
		= \sum_i \ket{\Psi_i},
\end{equation}
where the state vectors $\ket{\Psi_i}$, for different values of the index $i$, denote different classical realities, whereas the vector $\ket{\Psi}$ describes the complete quantum reality.

This formulation of the Everett's interpretation of quantum mechanics served as a basis for \emph{Quantum Concept of Consciousness} (QCC) proposed by the present author  \cite{Men00consEn, Men05consEn, MBM-QCC-2007-ENG, MBM-QCC-postcorrection-2007}.\footnote{It can also be called \emph{Extended Everett Concept} (EEC).} It is assumed in QCC that \emph{different classical realities are perceived by consciousness separately from each other}, so that, perceiving one of the classical realities, a human (and in fact any living creature) does not perceive the rest of them (forming quantum reality in their totality). This creates the impression (which is actually an illusion) that it is only a single classical reality that exists.

But in fact, the reality of our world is quantum, that is a set of alternative classical realities coexist in the quantum reality of the world. Therefore, in principle there can be some way for a person to gain access to ``other'' classical realities. According to QCC, this access occurs if the out-of-personal \emph{super-consciousness} is activated. This may occur when the personal consciousness is turned off, or even in parallel with it. 

It is clear that the information which may be collected from the state $\ket{\Psi}$ (which is available to super-consciousness) is immeasurably more complete and more accurate than the information contained in the subjectively perceived classical reality $\ket{\Psi_i}$ (accessible to consciousness).

According to quantum mechanics, evolution of the state vector $\ket{\Psi}$ (presenting the quantum reality) is linear. This evolution is described by a unitary evolution operator or linear Schr\"{o}dinger equation, and it is reversible. This actually means that all time moments are on the same footing with each other from the point of view of this state, i.e., from the viewpoint of the quantum reality. If the state $\ket{\Psi}$ is given at some time moment, then in principle it is known also at any other moment. Therefore, the \emph{``super-information''} is available for super-consciousness, which contains information about all possible classical realities of the world, not only in the present, but also in the past and in the future.

This, in particular, makes it possible for the super-consciousness such function as \emph{super-intuition}, or \emph{direct vision of the truth}. We mean the truth which is inaccessible to consciousness but exists in the quantum reality (in general, not necessarily exists in the present, may be in the past or in the future). A simple example of super-intuition is the scientific insight, when a scientist suggests, what solution of the problem of interest is correct (or what can be a way to solving this problem). The super-intuition, suggested by the super-consciousness, is always correct, even if the usual methods could logically lead to the correct solution only in the future.

\section{Super-consciousness as pure existence}
\label{sec:SuperconsciousnessAsExistence}

In Section~\ref{sec:QCCandSuper-intuition}, consciousness and super-consciousness were defined verbally. Consciousness is such a perception of the quantum world, that different classical realities of the quantum world (classical projections of this world) are separated from each other. Super-consciousness is the access to all (or at least many) classical realities forming the quantum reality. Let us see what quantum-mechanical formulas can be associated with these verbal definitions \cite{MBM-QCC-2013}.

\subsection{Separation of alternative classical realities as decoherence by an environment}
\label{sec:SeparationAsDecoherence}

By consciousness we understand the perception of the quantum world, in which its various classical projections (alternatives) are separated, that is the perception of one of them precludes the perception of others. What may correspond to this situation in quantum formalism? It is quite obvious that the separation of classical realities corresponds to \emph{decoherence}, i.e. to the transition of a pure state of a quantum system into a mixed state of this system due to interaction of the system with its environment. Decoherence leads to the description of the state of the system by a density matrix instead of a state vector \cite{Zeh1996inGiulini} (see also \cite{Men00bk}). This situation arises particularly in a quantum measurement, i.e. when the state of one quantum system (the measured system) is reflected in the state of another system (the meter).
 
Structure of the state of the complete system, which arises in the measurement, is described in Section~\ref{sec:QuMeasAndCoexistenceMacroDistinctStates}. In the context of the situation that we discuss now, the meter is the human brain, since the state of the outside world is reflected in the state of the brain. Denote the state of the brain (or, perhaps, some part of it) by $\ket{\Phi}$. The system which is measured (perceived) is all that is outside of the brain (or the mentioned part thereof), including the rest of the body and the outside world. The state of this external system will be denoted by $\ket{\psi}$. Then, according to Section \ref{sec:QuMeasAndCoexistenceMacroDistinctStates}, as a result of interaction between the two subsystems, the state of the complete system (of the quantum world as a whole) takes the form \footnote{Of course, we very much simplify the picture, but at the same time highlight the essential aspects of it.}
$$
\ket{\Psi}=\sum_i\ket{\Psi_i}=\sum_i c_i\ket{\psi_i}\ket{\Phi_i}
$$

If one is interested only in the state of the measuring instrument (in our case the brain), then one has to sum up over all the states of the measured system (the outside world). To do this correctly, one has to go over from the state vector $\ket{\Psi}$ to the corresponding density matrix $\proj{\Psi}$, and then take the partial trace of this matrix over the states $\ket{\psi_i}$ of the measured system (see, for example, \cite{Men00bk}). As a result, for the state of the meter (the brain), one obtains the density matrix \footnote{We assume that the states $\ket{\psi_i}$ are orthonormal.}
\begin{equation}
	\rho = \sum_i |c_i|^2 \proj{\Phi_i}
\end{equation}
This means that with probability $p_i=|c_i|^2$ the brain is in the state $\ket{\Phi_i}$, and this state of the brain means that the outside world is in the state $\ket{\psi_i}$ (with the same value of the index $i$). Transition, in describing the meter (brain), from the state vector $\ket{\Phi}$ to the density matrix $\rho$ is called decoherence of the meter under influence of its environment (the rest of the world).

Thus, we come to the description, in which the alternative realities $\ket{\Psi_i}=c_i\ket{\psi_i}\ket{\Phi_i}$ are incompatible with each other. This is the correct description of how the human consciousness works, that is how the state of his brain (or some part of the brain) reflects the state of the outside world (including the state of the rest of the body).\footnote{All these arguments refer to the perception of the quantum world by any living organism, although the words ``consciousness'' and ``brain'' should, in general case, be replaced with more appropriate terms.} Although with substantial simplifications, we have just derived this description in the framework of the ordinary quantum formalism.

\subsection{Super-consciousness as pure existence}
\label{sec:SuperconsciousnessAsPureExistence}

Super-consciousness is access to the quantum world without separating classical alternatives forming quantum reality of this world. This is an access to all classical alternatives simultaneously. It can be described by \Eq{Coexistence}. This formula represents the state of the whole world, not just the brain which perceives this world. Alternative classical images of the quantum world (classical projections of its state) are not separated. They do not exclude each other. Vice versa, they coexist.

This formula suggests some aspects of the phenomenon of super-consciousness, which are not fully reflected in the verbal formulation of QCC. The main issues that are originated from \Eq{Coexistence} are following:\footnote{on this, see also \cite{MBM-QCC-2013}}
\begin{enumerate}
	\item Super-consciousness is not a measurement (observation of one restricted system by another restricted system). Instead, super-consciousness is the state itself, and the state of the whole world. We can say that \emph{super-consciousness is nothing else than the very existence}.
	\item \emph{Super-consciousness is out-of-personal}, unlike consciousness, which is always personal. It is likely that the super-consciousness is just what Carl Jung called \emph{the collective unconscious.} 
	\item For a human in the state of super-consciousness there is no distinction between internal and external. In this state, the notion ``I'' becomes abstract and out-of personal, and this notion is identified with the notion ``world''. This is explicitly proclaimed in describing the state which an yogi achieves in the deep meditation.\footnote{Buddhists believe that their school of thought is not a religion but \emph{science of consciousness}, and experimental science, although the experiments they conduct, deal not with material devices, but with the processes in the mind.}

\end{enumerate}

It is easy to understand that the state ``I'' = ``the whole world'' (including the quantum features of the world) may also be naturally identified with the concept of God, which occurs in different religions (or rather, with what is common for all religions, what remains in any religion, if you omit the details imposed historically for clarity of the concept and for bringing a wider range of people to it). A man who falls into a deep meditation (who turned on the super-consciousness) becomes equal to God. For him, available become both 	super-intuition (direct vision of the truth) and even the management of his subjective reality (that is, those  alternative classical realities in which he would find himself after returning to the state of ordinary consciousness). Reflections on these issues may be found in the book \cite{MBMcons2010bk-EN}.

\section{Quantum information and correlations}
\label{sec:QuantumInformationAndCorrelations}

The notion of quantum information plays an important role in QCC. This leads to a problem in connection with theorem on the impossibility of quantum cloning, since the transfer of quantum information is impossible. However, substantial for QCC is not transfer, but only correlation between fragments of quantum information, so that the problem actually does not exist. 

\subsection{The problem of quantum information transfer}
\label{sec:ProblemTransferQuInfo}

Consider in detail the special case of super-intuition when insight arises from the fact that the information which the conscious mind can only meet in the future, becomes accessible to the super-consciousness and is used by the consciousness in the present. This situation is described by the mathematical operation called \emph{postcorrection} \cite{MBM-QCC-postcorrection-2007}. The essence of the phenomenon looks as if some information is transferred from the future into the present. This phenomenon can be called \emph{life-originated time machine}.

In connection with the assumption of this type of super-intuition, a problem, or an objection, arises \cite{Panov-QCC-2013}. The matter is that in quantum mechanics there is a \emph{theorem of the impossibility of quantum cloning}. According to this theorem, it is impossible to create such a device, which could turn an arbitrary state $\ket{\psi}$ of the given system into the state $\ket{\psi}\ket{\psi}$, i.e., could ``clone'' the state vector of the quantum system.

Theorem of the impossibility of quantum cloning is a consequence of the linear character of the evolution of any quantum system. Suppose we have created some system whose state is described by the vector $\ket{\Psi}=\ket{\psi}\ket{A}$, where the first factor presents a certain state of the subsystem which we want to clone, and the second --- an environment of this subsystem, including equipment for cloning. The fact that thus described device clones the state of the subsystem, means that for some time the entire system will change as follows:
$$
\ket{\psi}\ket{A} \rightarrow \ket{\psi}\ket{\psi}\ket{A'}
$$
The arrow indicates here the action of the evolution operator for a specified time.

Suppose that the system created by us operates correctly when the initial state of the subsystem has the form $\ket{\psi_1}$ or $\ket{\psi_2}$. This means that the following laws of evolution of these two states take place:
$$
\ket{\psi_1}\ket{A} \rightarrow \ket{\psi_1}\ket{\psi_1}\ket{A_1}, \quad 
\ket{\psi_2}\ket{A} \rightarrow \ket{\psi_2}\ket{\psi_2}\ket{A_2}
$$
Then, taking into account that the law of evolution is linear, we obtain for the initial state of the form $\ket{\psi_1}+\ket{\psi_2}$ the following evolution law:
$$
\Big(\ket{\psi_1}+\ket{\psi_2}\Big)\ket{A} \rightarrow 
\ket{\psi_1}\ket{\psi_1}\ket{A_1}+\ket{\psi_2}\ket{\psi_2}\ket{A_2}
$$
We see that even if $\ket{A_1}=\ket{A_2}=\ket{A''}$, the final state of the system is
$$
\Big(\ket{\psi_1}\ket{\psi_1}+\ket{\psi_2}\ket{\psi_2}\Big)\ket{A''}
$$
whereas the cloning law requires that the following state should arise
$$
\Big(\ket{\psi_1}+\ket{\psi_2}\Big)\Big(\ket{\psi_1}+\ket{\psi_2}\Big)\ket{A''}
= \Big(\ket{\psi_1}\ket{\psi_1}+\ket{\psi_2}\ket{\psi_2}
+\ket{\psi_1}\ket{\psi_2}+\ket{\psi_2}\ket{\psi_1}\Big)\ket{A''}
$$

So, even if the system clones any two states, it can not clone their sum. In other words, it is impossible to create a system that produces a clone of any state applied to the input. Quantum cloning of an arbitrary (unknown) state is impossible. 

We now turn to the conclusions that follow from this fact for quantum information.

Quantum information, in contrast to classical, is presented, in one way or another, by a state vector of some real quantum system. Actual information systems that convert quantum information, usually use qubits. Qubit (quantum bit) is a system that can be in one of two states $\ket{0}$ and $\ket{1}$ as well as in any linear combination of these states. This may be, for example, a photon, where the $\ket{0}$ and $\ket{1}$ may be two of its polarizations.

However, an arbitrary state $\ket{\psi}$ of an arbitrary quantum system also presents some quantum information. The difference from classical information is that the presentation of quantum information means real existence of a particular material system, the information carrier, whereas classical information is a text that can be written in any medium, rewritten to any other medium and multiplied in any number of copies. 

If a state vector can not be cloned, it is impossible to duplicate quantum information and then move any one of its copies, remaining the second unchanged. It is true that, with the aid of the procedure called quantum teleportation, quantum information can be transferred from one area to another. But after such transfer, quantum information in the source area is necessarily distorted, so that duplication (cloning) is not happening.

In context of QCC, this means that it is impossible to move arbitrary quantum information neither from the future into the present, nor generally from one classical alternative to the other. It would seem that it prohibits the use of information that occurs in the future, for correction (postcorrection) the present \cite{Panov-QCC-2013}.

This, however, is not entirely true. Careful analysis shows that the super-intuition requires not transfer from the future to the past quantum information, but correlation between some quantum information in the future with the quantum information in the present. Therefore, the key role in the phenomenon of super-intuition is played not by quantum information as such, but by \emph {correlation} of two fragments of quantum information, one of which is in the present, the second --- in the future. This is not forbidden by the theorem of impossibility of quantum cloning. 

\subsection{Correlation instead of the quantum information transfer}
\label{sec:CorrelationInsteadQuInfoTransfer}

Let us show that, when super-consciousness is working, not transfer of quantum information takes place, but correlation of various pieces of quantum information. Although this statement is common to various phenomena associated with the super-consciousness, it is convenient to demonstrate this assertion on a specific example. Consider the phenomenon of scientific insight that allows scientists to solve problems that require for its decision principally new approaches, new paradigms.

To make it possible to find an effective approach to solving difficult problems, in the brain of a scientist must be created a configuration that enables him to solve this problem. Let us characterize most general components of such a configuration and compare them to those in the brain of a scientist who already knows solution to the problem (this may be another scientist or even the same, but after a certain time when he at last found the solution of this problem).  

In the future, when this problem has to be solved, the brain of the scientist will contain a configuration comprising four components:
\begin{itemize}
	\item formulation of the given problem;
	\item assertion that the problem is solved;
	\item formulation $S$ of its decision;
	\item method $M$, which led to its decision.
\end{itemize}
At present, in the brain of the scientist who is working on the problem, some configuration also exists, but it has other components:
\begin{itemize}
	\item formulation of the problem;
	\item assertion that the problem has not been solved;
	\item formulations of possible solutions (say, $S_1$, $S_2$,~\dots);
	\item methods that may possibly lead to its solution (say, $M_1$, $M_2$,~\dots).	
\end{itemize}
In both cases, these components characterize the quantum state of the brain, i.e. corresponding quantum information.

By the theorem of the impossibility of quantum cloning, the quantum information from the future configuration can not be transferred to the present. However, super-consciousness, having access to all alternative classical realities at all times, can establish a correlation between intellectual configurations in the future and in the present. It turns out that this is enough (with sufficiently qualified scientist) to lead him to conjecture about the right decision or the right method, which leads to the solution of this problem.

The starting point that plays the role of a inquiry to super-consciousness, is the formulation of the problem that is already known for the scientist in the present. Having access to all the alternatives at all times, super-consciousness commits following steps which represent the minimum first and then deeper correlation between the quantum information in the present and in the future:
\begin{itemize}
	\item super-consciousness is looking for all alternative configurations in which this problem is formulated;
	\item selects among the found configurations those, in which this problem is solved.
\end{itemize}
This procedure selects, in the future, some alternative, which contains the quantum information, appropriately correlated with the information upon which the inquiry is based. Generally speaking, thus stands out not only a single alternative, but for simplicity we will only talk about one.

The fragment of the future quantum information  which is found in this way, simply by virtue of the standard scientific methodology, contains not only formulation of the given problem, but also formulation $S$ of its decision and the method $M$ which led to this decision. Super-consciousness is trying to find correlation between such aspects of quantum information in the future and in the present. This should be correlation between quantum information 1)~in the future (solution $S$ and method $M$, that led to the decision of the problem) and 2)~in the present (attempts at solving $S_1$, $S_2$,~\dots and provisional methods of solving the problem $M_1$ , $M_2$,~\dots).

It is possible that no correlation is found. Then the super-consciousness can not help to the scientist to find a solution. Perhaps the problem is formulated incorrectly and has no solution in this formulation, or the super-consciousness of this scientist is insufficiently effective and can not get access to the necessary alternative realities.

However, in some cases, super-consciousness discovers that solution $S$ coincides (is correlated) with one of those provisional solutions $S_a$, which this scientist was already thinking over. Then, returning to a state of normal consciousness, the scientist realizes that this coincidence (i.e., the correlation) allows him to discard the remaining provisional solutions and to stop on the decision $S_a$. Thus it turns out that he knows the solution, although can not prove that it really is the right solution. He was struck by a brilliant hunch, super-intuition has worked. 

It is possible that the complete solution $S$ can not be detected among options $S_1$, $S_2$,~\dots, but among those methods for solving $M_1$, $M_2$,~\dots, which scientist was considering, the method $M_b$ is detected that coincides with that method $M$, which in the future should lead to the correct decision. Then the scientist, not knowing the solution to the problem, nevertheless knows how to act to solve it. If his qualification is sufficient, he, after some (perhaps considerable) time, will come to the correct solution. In this case, the key role is also played by the super-intuition.

\subsection{Cognitive euphoria}
\label{sec:Cognitive euphoria}

In Section~\ref{sec:CorrelationInsteadQuInfoTransfer}, we made use of the expression ``a correlation is discovered'' (correlation between solutions or between methods of solving problems). But what does it mean? How super-consciousness ``tells'' the scientist (in a state of normal consciousness) that a correlation is found? Testimonies of the great scientists who have experienced the super-intuition, suggest that the role of the hints, or markers of detected correlations, is played by specific positive emotions.

In fact, in all cases, a brilliant hunch or super-intuition, is accompanied first, by full confidence that the hunch is correct and second, by strong positive emotions. This type of experiences may be called \emph{cognitive euphoria} (CE).

As a way of marking correlations, the cognitive euphoria plays a crucial role in the further work of the scientist on the problem. If the final formulation of the problem solution is found (suggested by super-consciousness), while it still is not proved on regular basis, cognitive euphoria helps to believe in this solution and not wasting time on false choices. If not the formulation of the solution but only the method which leads to it, is found, the cognitive euphoria is necessary in order to find the determination to work, applying this method, perhaps for a long time, and being confident in the correctness of the chosen path.

The scheme of action of super-consciousness that has been described above, allows to explain the non-trivial fact that the super-intuition comes only to those who were previously working on the problem seriously, using conventional rational methods, i.e. using ordinary consciousness. Without such preliminary study, the scientist would have at his disposal neither a set of possible solutions $S_1$, $S_2$,~\dots, nor possible methods $M_1$, $M_2$,~\dots of solving the problem.

If it were possible to move quantum information from the future to the present, such preliminary study of the problem would not be necessary. Ingenious conjecture could have been born without any effort, it would be available even for slacker. But since the transfer of  quantum information is not possible, and only correlation of fragments of quantum information is possible, \emph{preliminary attempts to solve the problem by rational methods is required}. Super-consciousness can help only to those who own, using ordinary consciousness and rational methods, is working hard to search for the truth.

\section{Concluding remarks}
\label{sec:Conclusion}

Above we have discussed how the phenomenon of scientific insights (a special case of super-intuition) can be explained in terms of Quantum Concept of Consciousness (QCC), or Extended Everett Concept. It is important that QCC can not be reduced to accounting quantum effects in the brain, as it is suggested in the work of Penrose and Hameroff \cite{HameroffPenrose2014}. According to QCC, brain (or special structures in the brain) plays the role of an interface between the body as a material system, and quantum reality as an expression of the specific conditions of existence in our world. According to QCC, quantum reality explains the relationship between consciousness and super-consciousness, one manifestation of which is the phenomenon of super-intuition. From the point of view of the quantum formalism, an essential role in this explanation is played by the notion of quantum information.

At first glance, super-intuition is connected with transferring information from the future into the present. However, it is not so. We have shown that emergence of super-intuition does not require transferring of quantum information, that would have been impossible without destroying the quantum information there, where it is extracted. What is really needed is to establish a correlation between the fragments of quantum information in different alternatives, and in particular at different times. 

However, establishing correlation between present and future means the access from the present to the information in the future. Transfer of quantum information does not occur, but the classical information emerges, which says that 1)~in some alternative in the future the problem in question receives its solution, and 2)~some approach to that solution, which is known already in the present, coincides with the approach that eventually will solve the problem. Thus obtained \emph{classical information} can be used in the state of normal consciousness. This information provides an important clue to solve the problem by rational methods, but does not replace this rational work.

We considered in detail the example of scientific insight. But in fact, the conclusion is common to all types of the phenomenon, which can be called \emph{super-knowledge}. In particular, this phenomenon easily explains jumps in the evolution of living beings which lead to new qualities, and generally the miracle of extreme efficiency of the evolution \cite{MBMcons2010bk-EN, Namiot-Biology-ENG}.

Let us finally mention that a deep analogy exists between super-intuition and \emph{principle of life} as the latter has been defined in the framework of QCC. This principle has been formulated in \cite{MBMcons2010bk-EN} in terms of Everett scenarios (chains of the alternatives, one alternative for each time moment). 

According to this principle, life is the set of the Everett scenarios which are favorable for living (the scenarios, that are the best to ensure survival). This is a global definition of life. However, the same may be defined locally, i.e. in the definite time moment. Let us take some moment, call it ``the present'' and try to formulate the principle of life as it looks in the present. This is done by the operation of \emph{postcorrection} \cite{MBM-QCC-postcorrection-2007} since this operation cuts off (in the present) those alternatives that generate unfavorable scenarios (not providing surviving in the future in the best possible way). 

What happens in the phenomenon of scientific insight, is quite analogous. A scientific insight does not mean getting the solution of the problem (the new paradigm) from the future. New ideas are born in the present. But the correlation with the future allows to cut off those alternatives which generate the scenarios, which do not lead to the correct solution. The remaining alternatives will generate the scenarios leading to the correct solution. \\

\textbf{\large Acknowledgement}\\[3mm]
The idea that specific features of quantum information play important role in Quantum Concept of Consciousness (known also as Extended Everett Concept), arose during discussions with L.V. Keldysh, whom I am truly thankful for this.

\newpage

\bibliography{BibTEXsourceFile}

\end{document}